\documentclass[10pt,letterpaper]{article}
\usepackage[top=0.85in,left=2.75in,footskip=0.75in,marginparwidth=2in]{geometry}

\usepackage[utf8]{inputenc}

\usepackage{cite}

\usepackage{nameref,hyperref}

\usepackage[right]{lineno}

\usepackage{microtype}
\DisableLigatures[f]{encoding = *, family = * }

\raggedright
\usepackage{changepage}

\usepackage[aboveskip=1pt,labelfont=bf,labelsep=period,singlelinecheck=off]{caption}

\makeatletter
\renewcommand{\@biblabel}[1]{\quad#1.}
\makeatother

\usepackage{lastpage,fancyhdr,graphicx}
\usepackage{epstopdf}
\fancyheadoffset[L]{2.25in}
\fancyfootoffset[L]{2.25in}

\usepackage{color}

\definecolor{Gray}{gray}{.25}

\usepackage{graphicx}

\begin{document}
\vspace*{0.2in}

\begin{flushleft}
{\LARGE
\textbf\newline{Ultrafast serrodyne optical frequency translator}
}
\newline
\\
Prannay Balla\textsuperscript{1,2,3},
Henrik Tünnermann\textsuperscript{1},
Sarper H. Salman\textsuperscript{1,2,3},
Mingqi Fan\textsuperscript{1},
Skirmantas Alisauskas\textsuperscript{1},
Ingmar Hartl\textsuperscript{1},
Christoph M. Heyl\textsuperscript{1,2,3,*}
\\
\bigskip
\bf{1} Deutsches Elektronen-Synchrotron DESY, Notkestr. 85, 22607 Hamburg, Germany
\\
\bf{2} Helmholtz-Institute Jena, Fr\"obelstieg 3, 07743 Jena, Germany

\bf{3} GSI Helmholtzzentrum für Schwerionenforschung GmbH, Planckstraße 1, 64291 Darmstadt, Germany
\\
\bigskip
* Corresponding author(s). E-mail(s): christoph.heyl@desy.de

\end{flushleft}

\textbf{The serrodyne principle enables shifting the frequency of an electromagnetic signal by applying a linear phase ramp in the time domain \cite{raymond1957}. This phenomenon has been exploited to frequency-shift signals in the radiofrequency (RF), microwave and optical region of the electromagnetic spectrum over ranges of up to a few GHz e.g. to analyse the Doppler shift of RF signals, for noise suppression and frequency stabilization \cite{winnall1997,chakam2002,mcdermitt2005,Houtz09,Johnson2010,Lee2021,Lauermann2016,Kohlhaas12}. Here, we employ this principle to shift the center frequency of high power femtosecond laser pulses over a range of several THz with the help of a nonlinear multi-pass cell.  We demonstrate our method experimentally by shifting the central wavelength of a state-of-the-art 75\,W frequency comb laser from 1030\,nm  to 1060\,nm and to 1000\,nm. Furthermore, we experimentally show that this wavelength shifting technique supports coherence characteristics at the few Hz-level while improving the temporal pulse quality. The technique is generally applicable to wide parameter ranges and different laser systems, enabling efficient wavelength conversion of high-power lasers to spectral regions beyond the gain bandwidth of available laser platforms.}


\vspace{0.5cm}

Ultrafast high-power lasers offer pulses with durations reaching the few-femtosecond range. Following the invention of chirped-pulse amplification (CPA), the last three decades are marked with major advances in ultrafast science and metrology as well as in strong-field physics and material sciences. An important part of this progress can be attributed to advances in ultrafast high peak and/or average power laser technology. High-power lasers opened a path to capture sub-femtosecond electron dynamics \cite{Corkum2007}, they have enabled great insights into the proton structure\cite{aldo2013} and the development of suitable tools for next generation chip manufacturing via short-wavelength nano-lithography \cite{Versolato_2022}. For many of the named innovations and applications, dedicated laser parameters are required including application-optimized wavelength, intensity, pulse length, beam properties and many more. In particular, the wavelength is a very important but yet not very flexible parameter. 

Today, Ytterbium- (Yb), Ti:Sapphire- (Ti:Sa) and since recently also Thulium-based laser platforms are commonly used for generating femtosecond pulses at high peak and/or average power. While these laser systems typically operate only at specific wavelengths defined by the bandwidth of the laser gain medium, modern laser technology provides a variety of options to build wavelength-tunable laser sources.  Wavelength-shifting approaches are routinely employed at low power levels using Stimulated Raman Scattering in optical fibers \cite{Stolen1972,Vanvincq_10}, soliton-shifting methods\cite{Mitschke86,Gordon86}, dispersive wave generation \cite{Wai86,Travers2019} and Raman-based shifting schemes in hollow-core fibers or capillaries \cite{Beetar2020-2,Tyumenev2022}. A wide spectral coverage and ultrashort pulses can also be provided by parametric frequency conversion employed e.g. in Optical Parametric Amplifiers \cite{Fattahi_14}. However, these methods typically suffer from low conversion efficiency or limited power handling capabilities. 

In contrast, pulse post-compression technology offers a route to reach ultra-short few-femtosecond pulse durations with high efficiency. In particular, when combined with Yb-based lasers, ultra-short pulses with kilowatts of average power \cite{Grebing2020} approaching the TW-peak power regime can be obtained \cite{Viotti22}. However, these and other high-peak power laser platforms lack a wavelength-tuning option.  Serrodyne frequency shifting methods, historically known from RF technology, can offer a solution to this problem. Translated from continuous wave signals to ultrashort laser pulses, the method can provide high efficiencies as well as compatibility to TW-peak and kW-average powers while supporting key characteristics required for precision metrology applications including phase coherence and carrier-envelope-offset frequency preservation.

The serrodyne principle states that, for a given signal, when a linear phase is applied in the time domain, the frequency of the signal shifts. Such a phase can be applied e.g. by electro-optic modulation \cite{chakam2002}. We instead use all-optical methods utilizing the Kerr-effect which can be employed to transfer a linear amplitude modulation in time into a linear phase ramp. 
A temporal saw-tooth pulse undergoing self-phase modulation (SPM) thus gets frequency-shifted. 
The magnitude and direction of the frequency shift is given by: 

\begin{equation}
 \Delta \omega = -\Phi_{NL} \frac{d I(t)}{d t},
 \label{eq:shift}
\end{equation}

where $I(t)$ is the time-dependent laser intensity and $\Phi_{NL}$ is the accumulated nonlinear phase, commonly known as $B$-integral. From Equation~\ref{eq:shift}, we can note that for a larger frequency shift, we can either increase the $B$-integral or the intensity gradient by creating a steeper pulse slope.
While the slope steepness is limited by the available spectral bandwidth, large $B$-integrals can be acquired by utilizing guiding concepts. However, for efficient SPM-based spectral shifting, the temporal pulse shape has to be maintained, which is typically not the case when SPM and linear dispersion are present.

The recent introduction of nonlinear multi-pass cells (MPCs)\cite{Schulte_16,Weitenberg_17} offers a solution to this challenge, providing the possibility to accomplish dispersion-balanced SPM supporting large $B$-integrals, thus enabling simple but very effective multi-THz frequency-shifting.

\begin{figure*}[!h]
\centering
\includegraphics[width=1\linewidth]{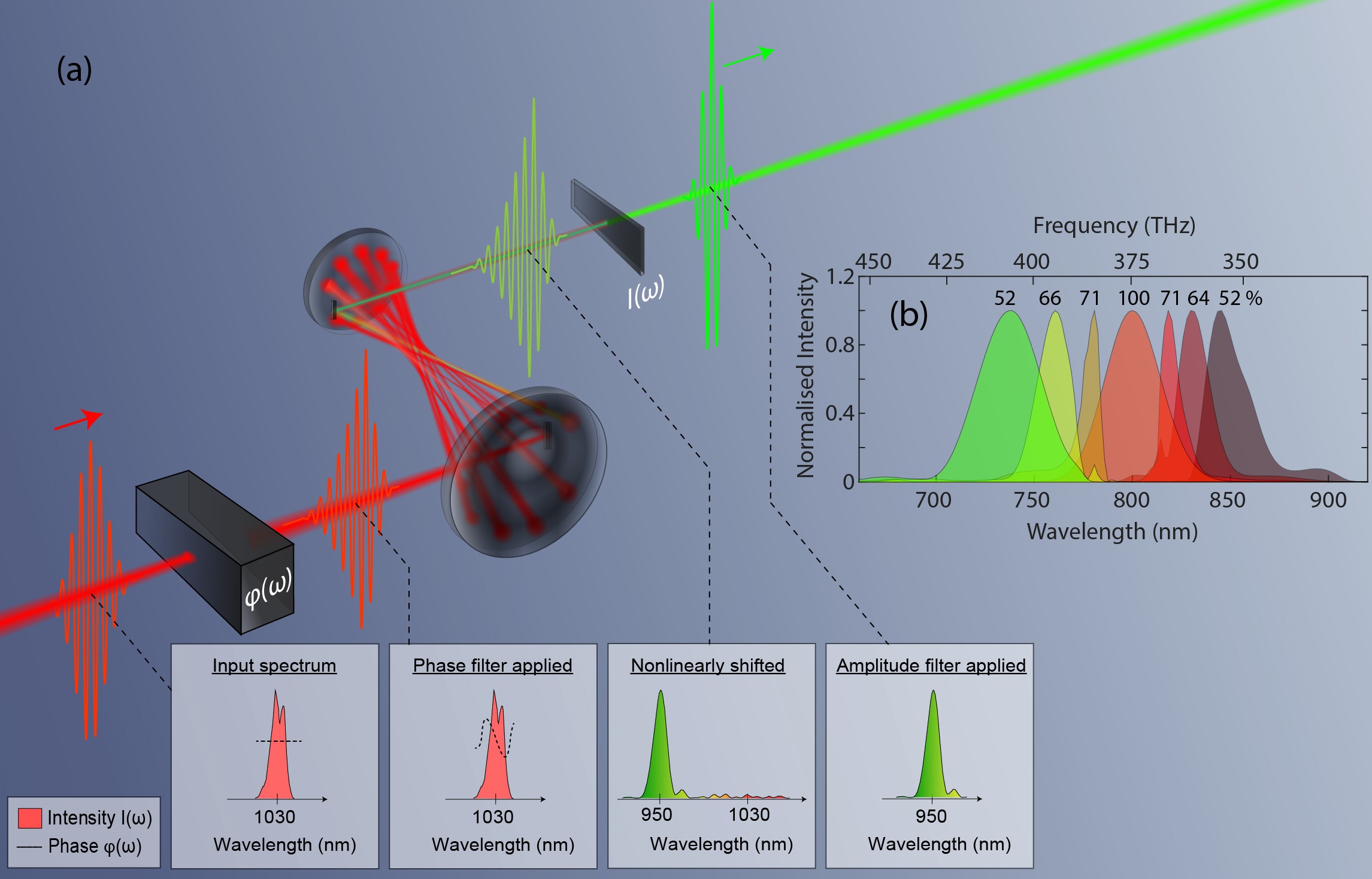}
\vspace{0.1cm}
\caption{{\bf Ultrafast optical serrodyne frequency shifting via self-phase modulation in a multi-pass cell.} \textbf{(a)} Illustration of the principle: Phase ($\varphi(\omega)$) and optionally amplitude shaping via a wave shaper enables the generation of a temporal saw-tooth pulse, which is sent though a dispersion-balanced nonlinear multi-pass cell yielding a frequency-shifted spectrum, which can be filtered using a dichroic filter ($I(\omega)$). \textbf{(b)} Simulation of the shifting process considering a 40\,fs input laser pulse centered at 800\,nm and a multi-pass cell made of commercially available broad-band matched-pair dielectric mirrors using two mirror bounces between consecutive passes through the focus. The displayed spectra and efficiencies take into account spectral shifting and filtering.}
\label{fig:principle}
\end{figure*}

Our frequency-shifting method is illustrated in Figure~\ref{fig:principle}(a). First, phase and optionally amplitude of an ultrashort laser pulse are shaped using a pulse shaper (PS). The PS can be implemented using e.g. a programmable spatial light modulator.
The laser pulse is then spectrally shifted in a second step using a dispersion-balanced MPC. Afterwards, the wavelength-shifted spectrum is separated from residual broadband wavelength components via a dichroic filter.

We numerically demonstrate our method considering a 40\,fs, Fourier-limited input pulse as e.g available from standard Ti:Sa laser systems. Using feed forward optimization routines (see Methods), we calculate an optimised phase providing maximum efficiency when shifting to a targeted output wavelength. Optimum efficiency is typically reached when the temporal pulse profile approaches a saw-tooth shape. The saw-tooth orientation thereby dictates the spectral shifting directions. Wavelength tuning is possible by simply changing the $B$-integral in the MPC and by optimizing the wave shaper settings. The resulting highly efficient wavelength shifting characteristics simulated considering realistic experimental conditions are shown in Fig.~\ref{fig:principle}(b). While our simulations already support a shifting range over 50\,THz with an efficiency larger than 50\%, higher efficiencies and larger frequency ranges can be expected with improved mirror characteristics.

\begin{figure*}[t!]
\centering
\includegraphics[width=1\linewidth]{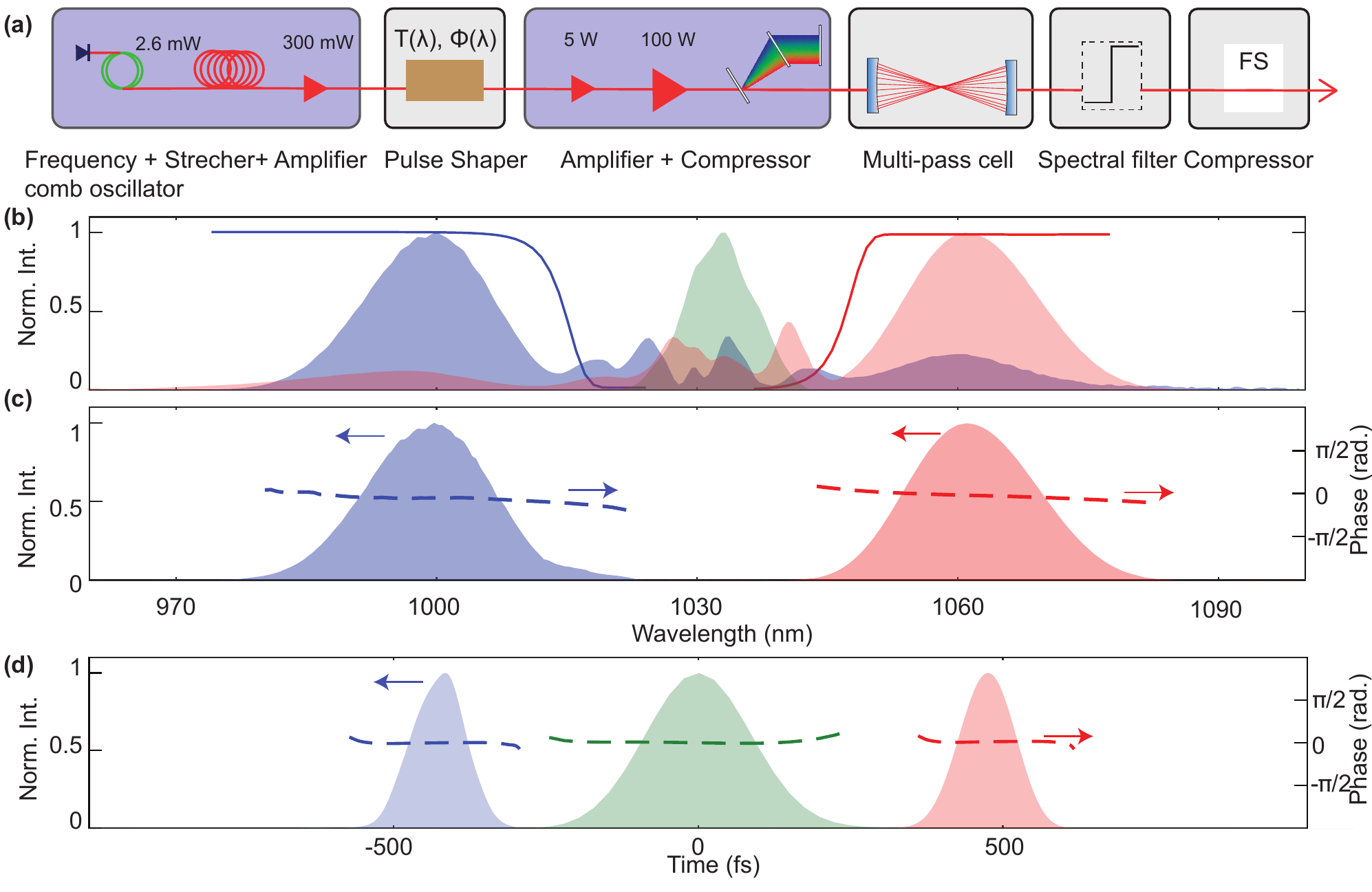}
\vspace{0.1cm}
\caption{\textbf{Setup and spectral shifting results.} \textbf{(a)} Schematic of the experimental setup including a high-power frequency comb laser (blue boxes) complemented with pulse shaper, MPC, spectral filter and fused-silica (FS) glass compressor for frequency shifting (gray boxes). \textbf{(b)} Measured spectrum at the MPC input (green) as well as at the output shifted to 999\,nm (blue) and to 1062\,nm (red) shown together with the transmission of the spectral filters (solid lines). \textbf{(c)} Corresponding measured spectra and retrieved phases of the output pulses after spectral filtering and compression. \textbf{(d)} Corresponding temporal intensity profiles and phases retrieved from second harmonic FROG measurements displayed together with the reconstructed MPC input laser pulse (green).}

\label{fig:setup}
\end{figure*}

To experimentally demonstrate our frequency shifting method, we employ a high-power Yb-based frequency comb system delivering 200\,fs pulses at 65\,MHz repetition rate. A schematic system layout is depicted in Fig.~\ref{fig:setup}. A pulse shaper introduced in the low-power section (340\,mW) of a three-stage amplifier chain utilizing CPA is used to introduce a phase optimized to produce a saw-tooth pulse after compression. At the CPA output, pulses at an average power of 76\,W are coupled into an MPC. The MPC consists of two dielectric concave mirrors with 100\,mm radius of curvature separated by a distance of 187\,mm. The dispersion of each mirror is approximately matched to 10.5\,mm UV fused silica within a 980-1080\,nm bandwidth, corresponding to multiple anti-reflection (AR) coated windows inserted into the MPC. The MPC is operated with 32 roundtrips yielding an overall transmission of 81\% corresponding to 62.2\,W at the MPC output (details see Methods section). The efficiency can be easily improved by reducing the number of optical windows or by improved AR coatings.

The input spectrum of the laser sent into the MPC as well as the corresponding output spectra after the MPC for optimized pulse shaper settings yielding spectral shifts to 999\,nm and to 1062\,nm are shown in Fig.~\ref{fig:setup}(b). The corresponding filtered spectra obatined using spectrally tunable dichroic mirrors are shown in Fig.~\ref{fig:setup}(c). After spectral filtering, we reach 66.7\% (41.5\,W) and 63\% (39.2\,W) of the optical power transmitted through the MPC centered at 1062\,nm and 999\,nm, respectively. We observed that the achievable efficiency depends critically on the MPC mirror performance and deteriorates with increasing nonlinear phase acquired in the CPA amplifier chain after the pulse shaper. A large nonlinear phase reduces the pulse-shaping performance at the CPA output.

Contrary to SPM-based spectral broadening in an MPC typically yielding a positively chirped pulse at the MPC output, we observe that the frequency-shifted pulses exhibit a small negative chirp. After spectral filtering, the wavelength-shifted pulses could be compressed to durations of 106\,fs (1062\,nm) and 92\,fs (999\,nm) by passing 119\,mm and 100\,nm fused silica, respectively. The temporal reconstruction from second harmonic frequency resolved optical gating (FROG) measurements of the laser output pulses at 1030\,nm as well as the spectrally shifted and compressed pulses are displyed in Fig.~\ref{fig:setup}(d), revealing excellent temporal pulse quality. In addition, we verify the spatial beam quality after spectral shifting. With an input beam quality parameter $M^2 = 1.2$x$1.4$, we obtain output beam parameters of $M^2 = 1.2$x$1.3$ at 1062\,nm and $1.3$x$1.6$ at 999\,nm.

\begin{figure}[t!]
\centering
\includegraphics[width=0.6\textwidth]{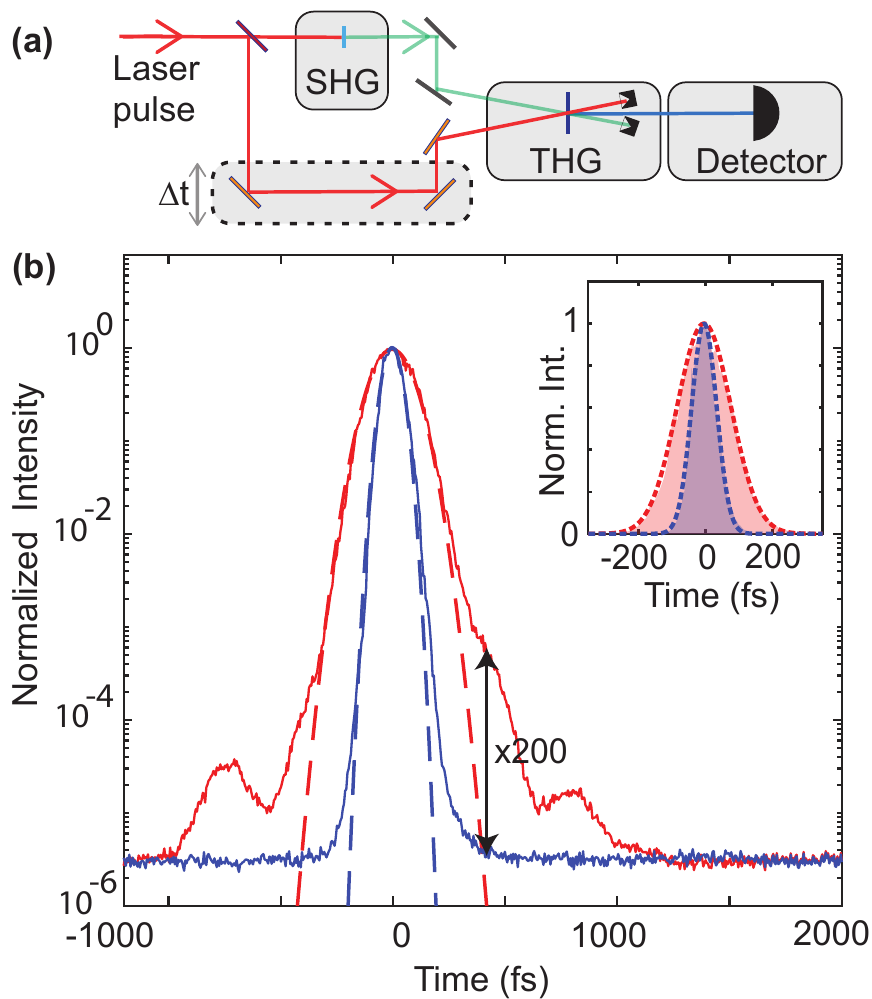}
\vspace{0.1cm}
\caption{\textbf{Temporal contrast characterization.} \textbf{(a)} Schematic of the third order auto-correlator used to measured the temporal contrast of the frequency-shifted pulses. \textbf{(b)} Measured auto-correlation signals (solid lines) for the laser output at 1030\,nm (red) and spectrally shifted and filtered pulse at 1062\,nm (blue). Gaussian curves (dashed lines) are fitted to the auto-correlation data. The inset in (b) shows the corresponding signals in linear scale.}
\label{fig:result2} 
\end{figure} 

Encouraged by the great temporal pulse quality revealed by the FROG measurements, we explore the temporal contrast of the frequency-shifted pulses. 
As in the RF domain, where frequency-shifting methods are employed for signal-to-noise reduction \cite{johnson88}, nonlinear frequency shifting of optical signals can enhance the temporal pulse contrast \cite{Buldt_17}. In order to verify this, we measure the third order auto-correlation of the pulse using the setup shown in Fig.~\ref{fig:result2}(a).  The intensity cross correlation between an input pulse and its second harmonic for an input center wavelength at 1033\,nm and the wavelength-shifted and spectrally filtered pulse at 1062\,nm is shown in Fig.~\ref{fig:result2}(b).  We find that the temporal pedestals are reduced by a factor of up to at least 200, limited by the dynamic range of our detector. In contrast to typical post-compressed pulses exhibiting temporal pre- and post-pulses \cite{Viotti2021}, the frequency-shifted pulses have a nearly perfect Gaussian shape over a large dynamic range.

\begin{figure}[t!]
\centering
\includegraphics[width=0.6\textwidth]{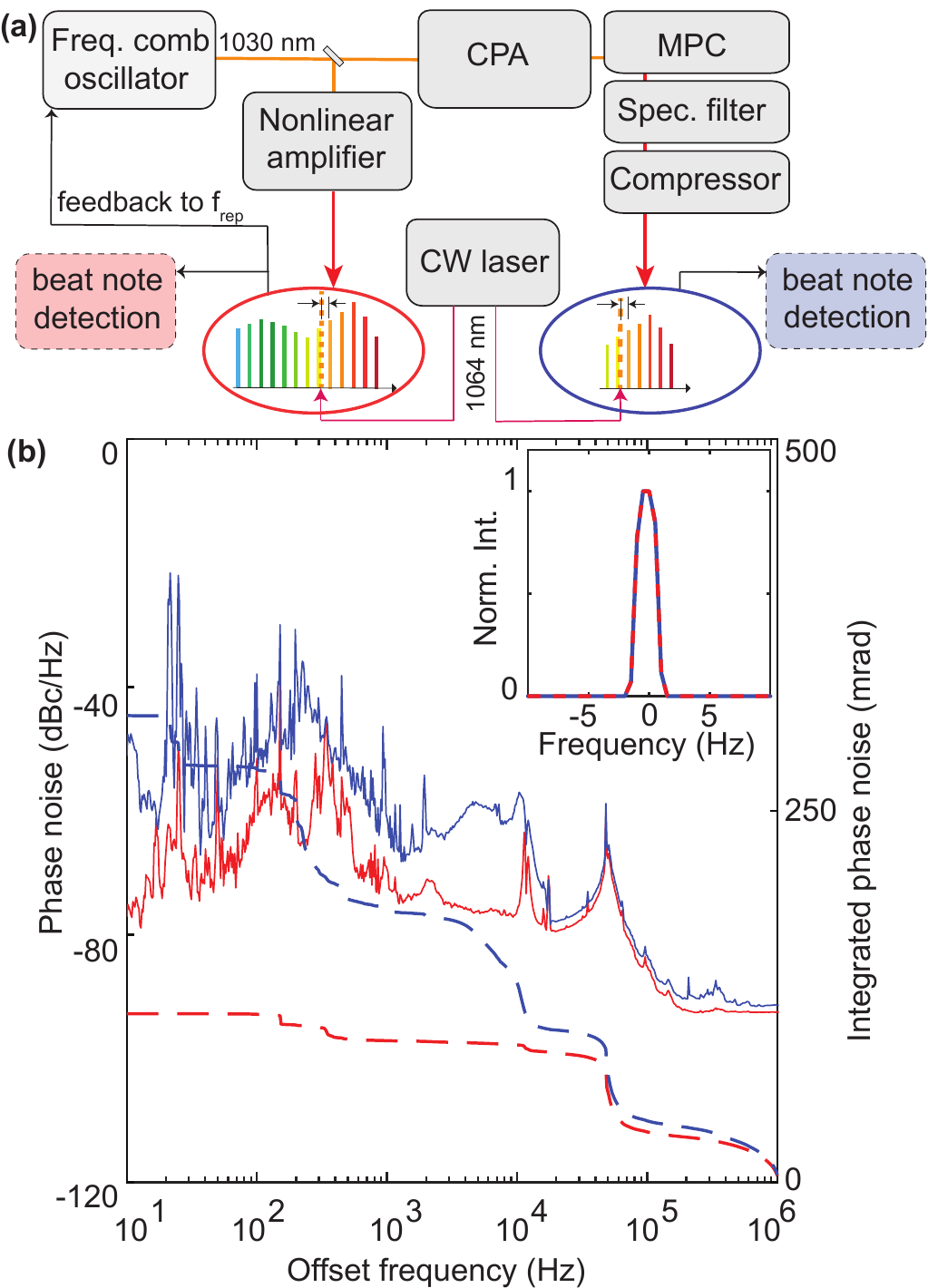}
\vspace{0.1cm}
\caption{\textbf{Coherence demonstration.} \textbf{(a)} Schematic of the setup used to characterize the coherence properties of the frequency-shifted high-power frequency-comb. A nonlinearly amplified portion of the oscillator output is used for stabilization to a CW reference laser at 1064\,nm. The same CW laser is used to produce a coherent beat note with the frequency-shifted laser. \textbf{(b)} Measured in-loop phase noise (red solid line), and beat note phase noise of the spectrally shifted laser with the CW laser (blue solid line) dislayed together with the integrated phase noise (dashed lines). The inset in (b) shows the corresponding beat notes measured on an RF spectrum analyser with a resolution bandwidth of 1.5\,Hz.}
\label{fig:result1}
\end{figure}

Another important property in particular for a frequency-comb lasers is its coherence. Coherence-properties of spectrally broadened laser pulses generated e.g. via super-continuum generation in highly nonlinear fibers, typically operated at an average power of a few mW, has been explored in detail \cite{Hartl2008}.  Here, we study the coherence characteristics of a spectrally shifted high power laser. We lock a nonlinearly amplified portion of our frequency comb oscillator to a stable continuous wave (CW) reference laser choosing a lock-offset frequency between both lasers of 20\,MHz. The CW laser has a linewidth around 1\,KHz and a wavelength of\,1064 nm. The schematic of the locking scheme is shown in Fig.~\ref{fig:result1}(a). The CW laser is further used to produce a heterodyne beat note with the frequency-shifted output after the MPC, centered at 1062\,nm. The measured beat note location at 20\,MHz resembles the chosen lock-offset, thus confirming preservation of the carrier-envelope offset frequency in the Serrodyne shifting process. Fig.~\ref{fig:result1}(b) displays the measured in-loop phase noise of the referenced laser (red) together and the phase-noise detected after frequency-shifting (blue). The inset shows the corresponding beat notes, indicating Hz-level linewidth support. We find that the integrated phase noise (10\,Hz to 1\,MHz) of the laser and spectrally shifted output amount to 113.5\,mrad and 314.1\,mrad, respectively, corresponding to about 99\% and 90\% of the power conatined in the carrier \cite{arimondo1993laser}. This result demonstrates excellent coherence properties of our method, setting an upper bound for a possible coherence degradation due to the shifting process.

In this work, we have introduced a versatile wavelength tuning method for ultrashort lasers supporting excellent temporal pulse quality and great coherence characteristics. We experimentally demonstrate our method shifting femtosecond laser pulses from 1030\,nm to 999\,nm and 1062 \,nm respectively, limited in shifting range mainly by the MPC mirror bandwidth. Using numerical simulations, we show that the method can be extended to larger spectral shifts, provided that sufficiently broadband mirrors are employed.

Our concept provides great prospects for the versatile implementation of wavelength-tunable laser sources covering large parameter ranges as it is based on nonlinear MPCs which have been demonstrated at pulse energies covering a few $\mu$J to 100 mJ, average powers reaching 1\,kW as well as pulse durations ranging from picoseconds to few optical cycles \cite{Viotti22}. 
In particular, when used with high-power Yb lasers and high-pulse energy MPCs \cite{Heyl_2022}, wavelength-tunable lasers with TW-scale peak powers and kilowatts of average power can come into reach \cite{Grebing2020,Stark2021}. The method thus promises an effective route for laser platforms providing high average powers, shot pulse durations and wavelength-tunability in a single unit, thus combining complementary advantages previously known from different laser architectures.
Serrodyne-frequency-shifted femtosecond lasers have thus the potential for boosting various spectroscopy applications ranging from remote sensing over frequency-comb spectroscopy to multi-photon microscopy. Furthermore, nonlinear spectroscopy methods and applications utilizing secondary laser-driven high harmonic sources as e.g. attosecond science could greatly benefit from continuously tunable extreme ultraviolet sources drive by wavelength-tunable lasers. Finally, applications demanding highest peak or average power lasers such as laser plasma acceleration\cite{Albert_2021} or semiconductor chip production \cite{Versolato_2022}, which were so far constraint to a single or very few operation points in the electromagnetic spectrum, will benefit from novel optimization opportunities.

\nolinenumbers





\pagebreak
\section*{Methods}\label{sec11}
\textbf{Laser and multi-pass cell.} The Yb laser used in this letter for frequency shifting consists of a NALM oscillator \cite{Ma2022} with an average output power of 12 \,mW, a pulse duration of 150\,fs and a repetition rate of 65\,Mhz. This oscillator is first amplified to 300\,mW in a fiber amplifier. Subsequently, a 165\,m polarization maintaining fiber stretches the pulses to 120\,ps. The stretched pulses are sent through a pulse shaper consisting of a programmable spatial light modulator. Afterwards, the pulses are amplified in a double-clad Yb doped fiber, followed by a rod-type amplifier delivering about 100\,W output power. Finally, using a grating compressor, the pulses are compressed to 205\,fs, with a Fourier limit of 190\,fs at a wavelength centered at 1033\,nm. At the input to the MPC, an average power of 76\,W is reached. The nonlinear phase accumulated in the fiber sections between wave shaper and laser output was estimated using numerical simulations resulting in 3.88\,rad.

\begin{figure}[h]
\centering
\includegraphics[width=0.6\linewidth]{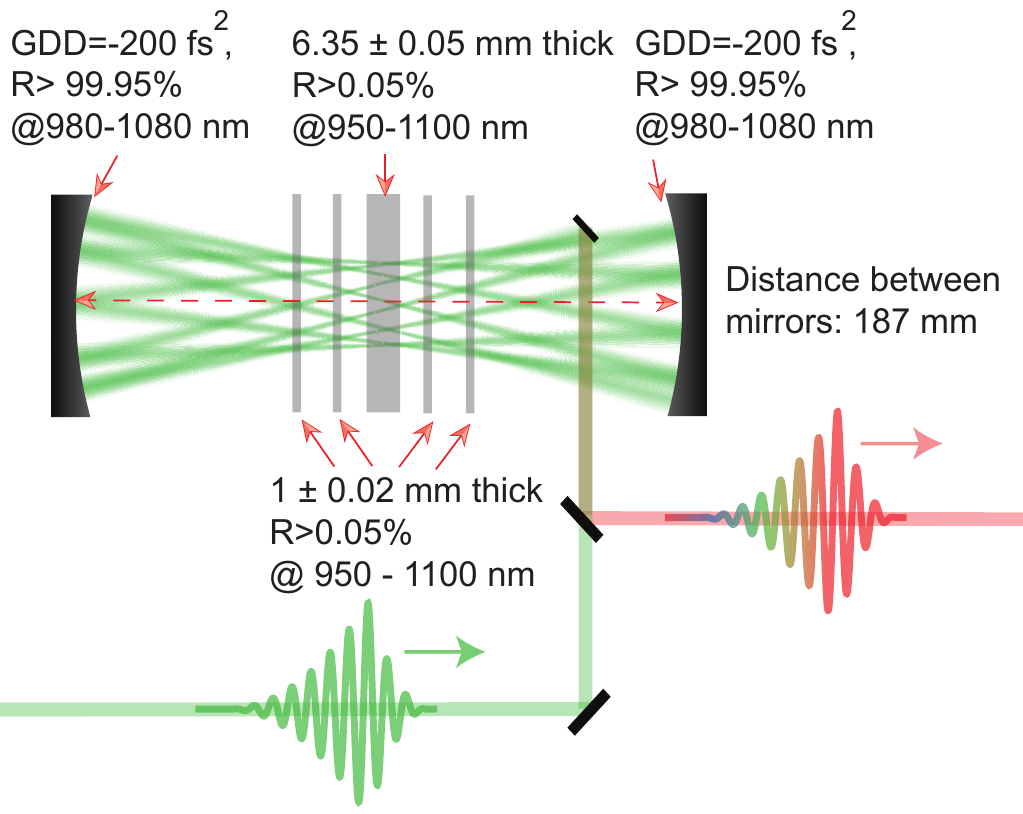}
\vspace{0.1cm}
\caption{\textbf{The multi-pass cell setup.}}

\label{fig:5}
\end{figure} 

A schematic of the compact MPC used in this experiment is shown in Fig.~\ref{fig:5}. The MPC consists of two concave mirrors with radii of curvature 100\,mm, separated by a distance of 187\,mm. The mirrors are coated with a dispersive coating of $-200 fs^2$ at 1030\,nm with group delay dispersion matched to 10.5\,mm fused silica. The nonlinear medium in this MPC consists of five anti-reflection coated solid fused silica windows. One window with a thickness 6.35\,mm is placed at the focus, four windows with a thickness of 1\,mm are placed at a distance of about 20\,mm and 40\,mm from the focus. The MPC accommodates 32 round trips. 
\\
\vspace{0.5cm}
\noindent
\textbf{Simulations and optimization.} The simulations for the spectral shifting results presented in Fig.\ref{fig:principle}(b), are performed by solving the Forward Maxwell Equation \cite{AC2011}.
The considered mirror characteristics are taken from commercially available dielectric mirrors providing dispersion properties matched to 2\,mm fused silica over a wide spectral range upon reflection at a mirror pair. In order to provide two mirror reflections in between consecutive passes through the nonlinear medium, an MPC consisting of four mirrors is required. The center wavelength of 800\,nm was chosen based on available broad-band coatings. The nonlinear phase accumulated per pass (about 1.7\,rad) lies within the range enabling multi-plate based MPC spectral broadening while supporting excellent spatial beam quality\cite{Seidel2022}. 

In order to optimize the phase required for spectral shifting, we employ a numerical library with support for automatic differentiation \cite{jax} and an optimization library originally developed to train neural networks (Optax \cite{optax}). We start the optimization process by launching a pulse with a spectral phase representing a good guess for a temporally asymmetric pulse form approaching a saw-tooth shape (transform limited pulse duration: 40~fs, Gaussian spectral amplitude) into the MPC. The phase is interatively optimized using Optax. This allows a relatively large parameter space to be optimized efficiently. A careful manual adaption of suitable target functions allows us to mitigate termination of the optimization process in local minima, a typical issue for gradient descent methods. 

\section*{Acknowledgements}
The authors acknowledge DESY (Hamburg, Germany) and Helmholtz-Institute Jena (Germany), members of the Helmholtz Association HGF, for support and/or the provision of experimental facilities. 

\section*{Author contributions}
P.B., H.T., S.S., M.F. and S.A. conducted the experiments, P.B. and H.T. performed the simulations, C.M.H conceived the initial idea, C.M.H. and I.H. supervised the project, all authors contributed to the writing of the manuscript.

\section*{Competing Interests}
The authors declare no competing interests.

\end{document}